\documentclass[twocolumn,aps,prb,showpacs]{revtex4-1}
\usepackage{epsfig,subfigure,amssymb,amscd,amsmath,graphicx,amsfonts}
\usepackage[bookmarks=false,linkcolor=blue,urlcolor=blue,colorlinks,citecolor=blue]{hyperref}
\usepackage[usenames,dvipsnames]{color}

\begin{document}

\title{Topological phases of parafermionic chains with symmetries}

\author{D. Meidan}
\affiliation{\mbox{Department of Physics, Ben-Gurion University of the Negev, BeÕer-Sheva 84105, Israel}}
\author{E. Berg}
\affiliation{\mbox{Department of Condensed Matter Physics, Weizmann Institute of Science, Rehovot 76100, Israel }}
\author{Ady Stern}
\affiliation{\mbox{Department of Condensed Matter Physics, Weizmann Institute of Science, Rehovot 76100, Israel }}
\begin{abstract}
We study the topological classification of parafermionic chains in the presence of a modified time reversal symmetry that satisfies ${\cal T}^2=1 $.
Such chains can be realized in one dimensional  structures embedded in fractionalized two dimensional states of matter, e.g. at the edges of a fractional quantum spin Hall system, where counter propagating modes  may be gapped either by back-scattering or by coupling to a superconductor.
In the absence of any additional symmetries, a chain of $\mathbb{Z}_m$ parafermions can belong to one of several distinct phases.
We find that when the modified time reversal symmetry is imposed, the classification becomes richer. If $m $ is odd, each of the  phases  splits into two  subclasses. We identify the symmetry protected phase as a Haldane phase that carries a KramersÕ doublet at each end.  When $m $ is even, each phase splits into four subclasses.
The origin of this split is in the emergent Majorana fermions associated with even values of $m$. We demonstrate the appearance of such emergent Majorana zero modes in a system where the constituents particles are either fermions or bosons.
\end{abstract}

\maketitle

\section{Introduction}
Much of the extensive research of topological states of matter in the last few years has focussed on topological classification of gapped systems \cite{Schnyder2008,Kitaev2009,Schnyder2010}. The question addressed in such classification is ``For a given dimension $d$ and symmetry $S$, what topologically distinct classes of Hamiltonians exist, where two Hamiltonians are topologically distinct if they cannot be deformed to one another without a closure of the energy gap?". This question was mostly studied for systems of non-interacting electrons and for systems which are adiabatically connected to non-interacting electrons~\cite{Fidkowski2010,Turner2011,Fidkowski2011a}. Here, we study this question in a setup in which interactions between electrons create two dimensional (2D) fractionalized states that are not adiabatically connected to systems of non-interacting electrons. We classify the gapped states that may be created in effectively one dimensional (1D) systems embedded in such systems. The 1D system may either exist at the edge of the host 2D phase, or along a line cutting through it.

One dimensional fermionic systems on a lattice may always be described in terms of coupled Majorana fermions. For non-interacting electrons it was realized that in the absence of any symmetry these systems fall in two topologically distinct phases~\cite{Hasan2010,Qi2011,Alicea2012,Stanescu2013}.
The topologically non-trivial phase is characterised by Majorana zero modes at its ends, which  are potentially useful for  protected quantum information processing. Interactions between electrons do not affect this classification \cite{Gangadharaiah2011,Fidkowski2011b,Stoudenmire2011,Sela2011}.
A richer structure emerges in one dimensional systems of the BDI class, composed of systems that conserve fermion parity and that are symmetric  to time reversal symmetry $\mathcal{T}$, with $\mathcal{T}^2=+1$. For these systems, in the absence of interactions the classifying group is $\mathbb{Z}$, while interactions reduce the classification to $\mathbb{Z}_8$ \cite{Fidkowski2010,Turner2011,Fidkowski2011a}. Hamiltonians of different classes differ by the number of Majorana zero modes at the systems' ends, and the reduction from $\mathbb{Z}$ to $\mathbb{Z}_8$ is a consequence of the fact that eight Majorana modes may be gapped by a local quartic interaction that does not violate the BDI symmetries. Similar reductions induced by interactions occur also for other symmetries and dimensions, and were classified in Refs. \onlinecite{Tang2012,Yao2013,Qi2013,Vishwanath2013,Gu2014,Wang2014,Senthil2015,Morimoto2015,Queiroz2016}.

In this work we go ``half a dimension higher". We study the classification of {\it one} dimensional systems that may occur only on edges of fractionalized {\it two} dimensional states of matter. 
An example of such a setup is the edge of a fractional quantum spin Hall system~\cite{Levin2009,Stern2016}, i.e., a 2D system in which electrons of one spin direction form a fractional quantum Hall state of filling $\nu$ while electrons of the other spin direction form a FQH state of filling $-\nu$. An edge structure that is equally good for our purpose forms also in more mundane systems - along a trench cut in a bulk of a fractional quantum Hall state, or along an edge of a double layer electron-hole systems where the electrons and the holes are at Landau filling fractions of equal magnitude and opposite sign. For concreteness, we will focus on the hypothetical construction of a fractional quantum spin Hall system.

Being interested in the classification of gapped systems, we note that such edges may be gapped either by (possibly spin-flipping) back-scattering between counter-propagating modes or by coupling of the counter-propagating modes to (either singlet or spin-polarized) superconductor. At interfaces between two semi-infinite regions of these two types of gapping mechanisms topologically protected zero modes occur. When the gapped edge modes are of integer quantum Hall states the zero modes are Fu-Kane Majorana states \cite{Fu2008}. When the gapped edges are of fractional quantum Hall states, the zero modes are ``fractionalized Majorana modes," \cite{Linder2012,Clarke2013,Cheng2012,Barkeshli2013} frequently referred to as ``parafermionic zero modes". These zero modes are localized unitary operators $\xi$ that commute with the Hamiltonian and satisfy $\xi^m=1$. The value of $m$ depends on the case at hand and will be discussed in detail below.

\begin{figure}[t]
\includegraphics[width=0.35 \textwidth]{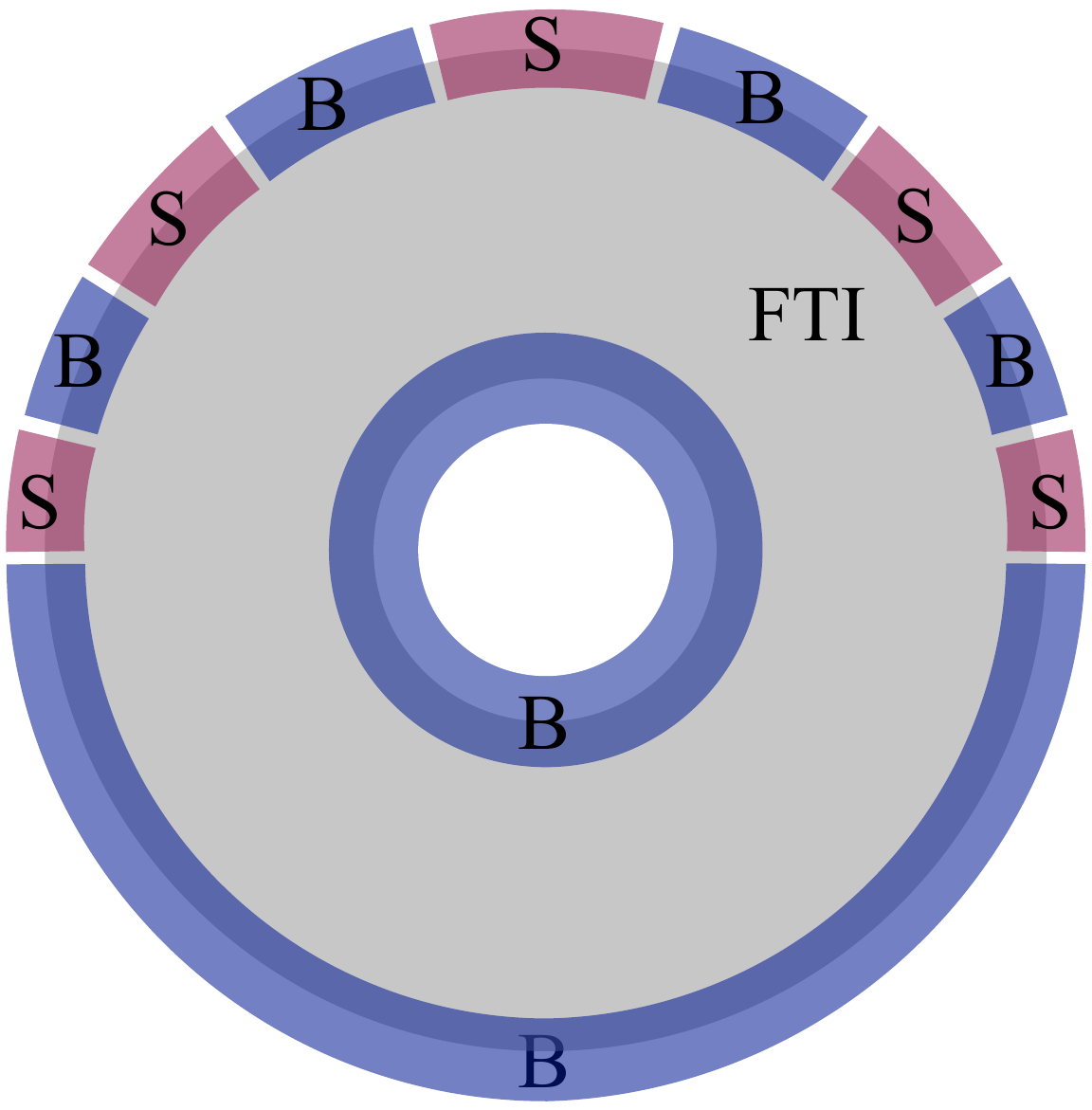}
\caption{A physical realization of a one dimensional parafermionic chain on the edge of a  two dimensional fractional topological insulator (FTI). The  outer edge of  the FTI is gapped by   alternating  superconducting regions (S) and back-scattering regions (B). At interfaces between these regions, parafermionic modes occur. The long B stretch on the outer edge can be thought of as the vacuum, leaving  the rest of the outer edge as  an open  1D system. }
\label{system}
\end{figure}
A central example is schematically depicted in Fig.~\ref{system}. The edge of a fractional quantum spin Hall state is gapped by alternating back-scattering (B) and superconducting (S)  regions. At interfaces between these regions, parafermionic modes occur. For infinite separation between the interfaces the ground state is multiply degenerate. The regions are however not infinite, and therefore parafermions of different interfaces couple. The coupling removes the degeneracy.
We consider a long edge where half of the edge is gapped by a long stretch of normal back-scattering $B$ and the other half is gapped by short alternating  superconducting regions ($S$) and back-scattering regions ($B$). The long $B$ stretch can be thought of as the vacuum, leaving the rest of the edge as a finite 1D system. We allow all local couplings within this system and classify the resulting gapped phases.

To set up the problem, we focus first on the case of two counter-propagating edge modes of $\nu=\pm 1/m$, with $p$ odd and prime. The parafermionic zero modes $\chi_{is}$ (with $i$ enumerating the interface, and $s=\pm 1$ is the spin index) are unitary operators that satisfy $\chi_{is}^{2m}=1$, and obey parafermions statistics \cite{Fradkin1980,Fendley2012},
\begin{align}
\chi_{is}\chi_{js'}=
\chi_{js'}\chi_{is}  e^{is \pi/m}
\label{zeromodes1overm}
\end{align}
for $ i<j$.
In the absence of any additional symmetries, the parafermionic chain is described by the Hamiltonian:
\begin{align}
H= \sum_{ij,s} t_{ij}^s\chi_{is}^\dag\chi_{js}+ \sum_{ijkl,ss'} V_{ijkl}^{ss'}\chi_{is}^\dag\chi_{js'}^\dag\chi_{ks'}\chi_{ls}+...
\label{ham1overm}
\end{align}
We assume that the Hamiltonian is \emph{local}, i.e. that the coupling constants decay exponentially with distance. 
The question we ask is ``what are the gapped phases that result from the Hamiltonian (\ref{ham1overm}) for particular symmetry constraints?". Note that the case $m=1$ is nothing but the one dimensional system of coupled Majorana modes. For general odd $m$, the elements $(\chi_{is})^n$ (with $n= 1, \dots, 2m$) form a representation of the $\mathbb{Z}_{2m}$ group, which may be written as a product of the two subgroups $\mathbb{Z}_2\times \mathbb{Z}_m$. As a consequence, each interface hosts a $\mathbb{Z}_2$ parafermion, which is a Majorana fermion, and a $\mathbb{Z}_m$ parafermion.

In our analysis below we use the $\nu=\pm 1/m$ example to introduce various concepts, but our discussion goes beyond that case, to more complicated abelian fractional quantum spin Hall states.

The structure of the paper is the following: In Sec. \ref{results} we summarize our main results with a focus on the underlying physical picture. The derivation based on the fractionalization of symmetries  is presented in Sec. \ref{fractionalization of symm}. The gapped phases identified  in Sec. \ref{fractionalization of symm} are then explicitly constructed in Sec. \ref{construction}, and  Sec. \ref{conclusion} contains concluding remarks.

\section{Physical picture and summary of results}\label{results}
\subsection{A  review of the classification of  parafermionic chains without symmetries}
The topological classification of Hamiltonians of coupled $\mathbb{Z}_N$ parafermions in the absence of any additional symmetries was obtained in Refs.~ \onlinecite{Motruk2013,Bondesan2013,Alexandradinata2016}. The $\mathbb{Z}_N$ parafermions may be deconstructed according to the decomposition of the cyclic group $\mathbb{Z}_N=\mathbb{Z}_{p_1^{r_1}}\times \dots \times \mathbb{Z}_{p_j^{r_j}}$, where $N=\prod_{i=1}^j p_i^{r_i}$ is the decomposition of $N$ into primes $p_i$.
The $\mathbb{Z}_N$ parafermion then breaks into $j$ different parafermions that act on different Hilbert spaces.
 In the absence of any additional symmetry a chain of $\mathbb{Z}_N$ parafermions can be in several distinct classes corresponding to the presence or absence of $\mathbb{Z}_{p_i^{r_i}} $ parafermions at its end. In addition, if
 $r_i>1$,
groups of parafermions whose self statistics is bosonic may condense throughout the entire edge. When this happens, the uncondensed parafermions  form a smaller cyclic group. The overall number of distinct phases is then $N_{\mathrm{phases}} =  (r_1+1) \dots (r_l+1)$, equal to the number of subgroups of $\mathbb{Z}_N$.

Below we will extend this classification to a chain of $\mathbb{Z}_N $ parafermions with an effective time reversal symmetry.  Before we do so, we will briefly describe the origin of such an emergent time reversal symmetry in our fractionalized system.

\subsection{Time reversal symmetry in the fractionalized case}
The fractional quantum spin Hall insulator is symmetric to time reversal $\mathcal{T}$ with$\mathcal{T}^2=(-1)^{n_F}$, where $n_F$ is the number of electrons in the system. Time reversal transforms $\psi^\dagger_\uparrow \rightarrow\psi^\dagger_\downarrow$ and $\psi^\dagger_\downarrow \rightarrow -\psi^\dagger_\uparrow$ (with the $\psi^\dagger$ being electron creation operators, and the subscripts denoting spin directions). For $\nu=1/m$, the backscattering term necessarily violates time reversal symmetry, since it corresponds to a Zeeman coupling to a magnetic field. However, when the magnetic field is purely in the $x-y$ plane the Zeeman coupling is of the form $\lambda\psi^\dagger_\uparrow\psi_\downarrow +h.c.$ and is symmetric under a modified time reversal symmetry ${\cal T}$, which is given by $\psi^\dagger_\uparrow \rightarrow\psi^\dagger_\downarrow$ and $\psi^\dagger_\downarrow \rightarrow \psi^\dagger_\uparrow$.  Physically, this symmetry operation corresponds to time reversal followed by a rotation by $\pi$ around the $z$ axis. Remarkably, ${\cal T}^2=1$. Thus, under these conditions the problem we consider may be regarded as a fracionalized generalization of the one dimensional systems of class BDI.

For more complicated fractional quantum spin Hall states of $\pm \nu$, the edge may be gapped in a way that does not break the original time reversal symmetry with $\mathcal{T}^2=-1$. This is allowed when the ratio $\nu/e^*$ is an even number, with $e^*$ being the charge of the smallest charge quasi-particle \cite{Levin2009}.
In that case
 the entire system belongs to the one dimensional DIII class.

\subsection{A short review of the non-fractionalized case ($m=1$)\label{review1dbd1}}
The $m=1$ case is topologically equivalent to a one-dimensional spinless fermionic system on a lattice. In the absence of symmetry, a pair of Majorana zero modes in close proximity may be gapped by their mutual coupling. Thus, there are two phases that are topologically distinct from one another, with zero or one Majorana mode at each end.

In the BDI class, with time reversal symmetry that satisfies ${\cal T}^2=1$, each of the two phases splits into four. This is so since a coupling of two neighboring Majorana fermions may violate time reversal symmetry, but a quartic coupling generally does not. With the most general coupling allowed by the symmetry, the class with zero Majorana modes splits into four subclasses: One trivial subclass with no end modes ($k=0$); two subclasses denoted by $k=\pm 2$ with two Majorana modes at each end, for which the fermion parity at each end is flipped by time reversal symmetry (a property for which they are termed anomalous);   and one subclass with four Majorana modes at each end (denoted by $k=4$). The $k=4$ subclass is topologically equivalent  \cite{Meidan2014} to a Haldane spin-$1$ chain \cite{Haldane1983a,Haldane1983b}, and carries a Kramers' doublet at each end.

The class with a single Majorana end mode splits into four subclasses as well: two subclasses with one mode at each end ($k=\pm 1$) and two subclasses with three Majorana modes at each end ($k=\pm 3$).
 Here $k $ and $ -k$ are inverse of each other,  in the sense that when combined, the phase and its inverse form a trivial phase. The phase and its inverse have similar physical properties.

 Altogether, then, Hamiltonians of a chain of $\mathbb{Z}_2$ parafermions are classified by a $\mathbb{Z}_2$ group in the absence of symmetries, and by $\mathbb{Z}_8$ group in the BDI case. There are several ways by which this $\mathbb{Z}_8$ classification may be derived \cite{Fidkowski2010,Turner2011,Fidkowski2011a,Meidan2014}. Most useful to our context is the derivation based on the fractionalization of symmetries \cite{Turner2011}. It is reviewed and generalized in Section \ref{fractionalization of symm}.

In the following section we generalize this classification to a chain of $\mathbb{Z}_N$ parafermions. We note, however, that unlike its  one dimensional analog, there is no obvious way to define an addition of gapped phases for a system  confined to the edge of a fractionalized bulk. In particular the notion of stacking is  ill-defined. Hence, in our analysis  below we restrict ourselves to counting the number of distinct phases and do not address the group structure.

\subsection{Classification of parafermionic chains with time reversal symmetry: the general principle}
The results derived in Sec. \ref{fractionalization of symm} can be summarized in a very concise way:  In the absence of any symmetry, a chain of $\mathbb{Z}_N $ parafermions falls into several topological classes, corresponding to the number of subgroups of $\mathbb{Z}_N$ (equal to the number of distinct divisors of $N$).
With modified time reversal symmetry ${\cal T}^2=1$ (corresponding to symmetry class BDI), each of these phases splits into four distinct subclasses if $N $ is even, and to two subclasses if $N $ is odd.

When the parafermionic chain is symmetric to time reversal symmetry $\mathcal{T}^2=-1 $, the system  belongs to DIII class. In this case the zero modes decompose into interface operators which are Kramers' pairs of Majorana modes and domain operators which are fractional. The  $\mathcal{T}^2=-1 $ time reversal symmetry has no effect on the fractional part. The parafermionic chain can be in several distinct classes corresponding to the presence or absence of  Kramers' pairs of Majoranas or of $\mathbb{Z}_{p_j^{r_j}} $ parafermions zero modes at its end.

Once the  general classification procedure is laid out, the problem reduces to the task of identifying  the interface $\mathbb{Z}_N$ parafermions for the physical problem at hand. In subsection~\ref{subs1}  we will introduce general considerations regarding zero modes in fractionalized edges. Then, in subsection (\ref{subs3}) we give examples for fermionic (\ref{subs31}) and bosonic (\ref{subs32}) systems.

\subsection{Ground state degeneracy and composition of zero modes \label{subs1}}

We consider the system depicted in Fig.~\ref{system}, where the edge of a fractional quantum spin Hall state is gapped by alternating back-scattering ($B$) and superconducting ($S$)  regions.
As we now show, the parafermions at the interfaces between  $B$ and $S$ regions may generally be understood as a combination of interface Majorana operators with domain operators that originate from the $B$ and the $S$ regions. The topological properties of the latter are determined by the 2D bulk.

When the $S$ and $B$ segments are very large, the ground state is multiply degenerate. The degeneracy of each gapped segment determines the composition  of the parafermionic zero modes.
Coupling to a superconductor breaks charge conservation to a lower symmetry group.
If the constituent particles are fermions, fermion parity is conserved  in addition to the fractional part of the charge. Thus for $\nu=\pm 1/m$ the resulting degeneracy is $2m$ per segment.  In this case, the zero modes are  composed of a $\mathbb{Z}_2$ parafermion (Majorana fermion) $\gamma $ and a $\mathbb{Z}_m$ parafermion which we denote as $\eta $: $\chi_i = \gamma_i\eta_i $.
Conversely, if the constituent particles are bosons, charge conservation is fully broken since bosons can be created and annihilated at the domain wall,  and fractional charge is conserved up to a charge of a single boson. The zero modes in these phases do not have a fermionic part, and  $\chi_i = \eta_i$.

\begin{figure}[h]
\includegraphics[width=0.35 \textwidth]{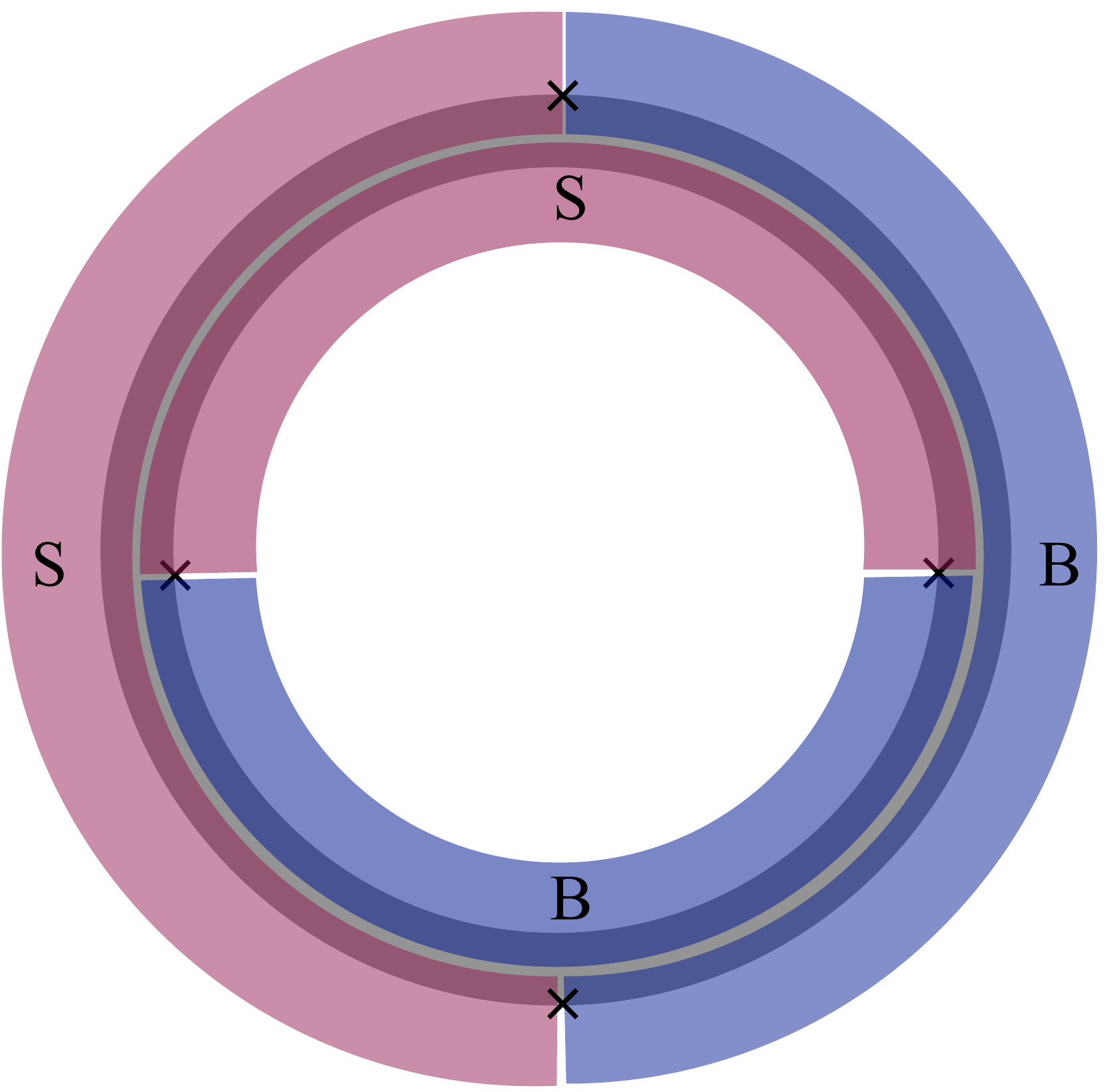}
\caption{To highlight the distinction between   interface zero modes and domain zero modes we consider an annulus slab of a fractional topological insulator where  the inner and outer edges of the annulus are covered by alternating S and B regions, such that the interfaces that pin parafermionic zero modes on the two edges are kept far from one another. In the thin annulus limit, the system becomes one dimensional and consequently the remaining   zero modes are Majorana fermions. Since the interfaces between the gapped sections on the inner and outer edges of the annulus remain  macroscopically distant even in the thin annulus limit, the removal of the non-Majorana zero modes is  a result of  domain operators. }
\label{thin_annulus}
\end{figure}

The distinction between interface zero modes and domain zero modes may be highlighted by considering an annulus slab of a fractional topological insulator where  the inner and outer edge of the annulus are covered by alternating $S$ and $B$ regions, such that the interfaces that pin parafermionic zero modes on the two edges are kept far from one another, see Fig. \ref{thin_annulus}.
In the thin annulus limit the system becomes one dimensional, and must conform to the classification of one dimensional systems. As such, the ground state degeneracy must be reduced and the only zero modes must be composed of Majorana fermions. \cite{Turner2011,Fidkowski2011a}
Since the interfaces between the gapped sections on the inner and outer edges of the annulus stay macroscopically distant even when the annulus becomes thin, there must be a way for the ground states to couple through domain operators rather than interface operators. Put differently,  the removal of the non-Majorana ground state degeneracy is a result of the proximity of the gapped regions to one another, and there must be operators that act within the gapped regions and have matrix elements between ground states. Those ``domain operators'' would be responsible for the removal of the ground state degeneracy that characterizes the annulus when both edges are entirely gapped by a single mechanism. When the gapping is by back-scattering, the domain operators transfer a $\pm e^*$ dipole between gapped regions, while when the gapping is by a superconductor, the transferred object has a charge $e^*$ in each spin direction. The interface zero modes may then be expressed as composed of Majorana modes which are localized at the interfaces and remain stable in the thin annulus limit,  and two domian operators, one in the $S$ region and one in the $B$ region. The Majorana operators commute with the domain operators (as with any fractional excitation.) \cite{Note1}. Hence, the two operators act in disconnected Hilbert spaces, and thus we may consider them separately.

The low energy subspace of the gapped edge may be described in terms of the charge degree of freedom of the S regions or the spin degree of freedom of the B regions. The charge on each superconducting region may take the values $q\mod 2=j/m$ with $j=0,1,\dots,(2m-1)$. The interface Majorana operators change this charge by one. The domain operators acting within the superconductor change it by $2/m$. The interface operator $\eta$ is a combination of the latter two, and may be defined as an operators that adds two quasi-particles of one spin direction to an interface. As such, it changes the charge on the $S$ region and the spin in the $B$ region that neighbor with the interface on which it operates. For odd $m$, there is only one combination of Majorana and domain operators for every charge $\Delta j$ added to the superconductor. The same description holds also for the spins in the B region. This is in consistency with the observation that the $\mathbb{Z}_{2m}=\mathbb{Z}_2\times \mathbb{Z}_m$ for odd $m$.

Although this discussion was presented for $\nu=\pm 1/m$, some parts of it are general. For all the classification questions that we defined here, the first step is to identify the interface parafermionic zero modes. Once those are identified,
 the edge is described as a chain of coupled parafermion zero modes.

\subsection{Examples  \label{subs3}}
\subsubsection{Fermionic systems}\label{subs31}

Let us first consider the case of a prime $m$, with $\nu=\pm 1/3$ being the prominent example.
As explained above, for $\nu=1/m$ the edge system we consider may be mapped to two disconnected subsystems - a one dimensional chain of coupled Majorana modes, and a one dimensional chain of coupled $\mathbb{Z}_m$ parafermionic modes. The classification of the Majorana chain is reviewed in subsection (\ref{review1dbd1}). In the absence of additional symmetries except fermion parity conservation, it has two topological classes, with a $\mathbb{Z}_2$ structure.
The two classes have obvious analogs for the parafermions: a trivial phase is created if the S regions are all shorter than the B ones. Then, the edge becomes an insulator. The topological phase is created if the $S$ regions are longer than the $B$ regions. Then the system becomes half-superconducting and half insulating, with two parafermions at the interfaces \cite{Motruk2013}. Altogether, then, the gapped edge falls into one of  four gapped phases, depending on the presence or absence of a $\mathbb{Z}_2  $ Majorana  and of a $\mathbb{Z}_m$ parafermionic modes  at the interface with the insulating (B) region.

When the modified time reversal symmetry is imposed, the classification becomes richer, and  each of  topological phases is split into four due to the appearance of two symmetry protected phases, the Haldane phase ($k=4$), and the anomalous phase ($k=2$). The combined system of Majorana modes and $ \mathbb{Z}_m$ parafermions then classifies to  16 gapped phases.

While this exhausts all topological phases of $\mathbb{Z}_{2}\times \mathbb{Z}_{m}$ parafermions, it is instructive to note that the Haldane phase can be constructed even in the absence of the $ \mathbb{Z}_2$ Majorana fermions. For this purpose we consider coupling the $ \mathbb{Z}_2$ Majorana fermions across the superconducting domains. This lifts the degeneracy associated with fermion parity, leaving each S domain with a $m$ fold degenerate ground state labelled by its fractional charge.
Since the $ \mathbb{Z}_m$ parafermionic operators change that spin by $2/m$, they span a Hilbert space of $m$ states per B region. Thus, each B region may be regarded as a spin-$(m-1)/2$. With $p$ being odd, this is an integer fictitious-spin. The effect of time reversal symmetry on the fictitious spin is the same as its effect on a real integer spin. Thus, the problem is mapped onto a chain of integer spins whose coupling is time reversal symmetric. Such a system can have a topological Haldane phase that  has Kramers' doublet on each of its two ends.

Interestingly, there is no $\mathbb{Z}_m$ analog to the anomalous $k=\pm 2$ subclasses of the $\mathbb{Z}_2$ case. This can be understood by noting that in the anomalous $k=\pm 2$ of the Majorana chain time reversal flips the fermion parity. As must be the case, two operations of time reversal bring the system back to its original state. For $\mathbb{Z}_m$ parafermions, fermion parity is replaced by the fractional part of the charge, which has $m$ possible values. For odd $m$, it is then impossible to have a $\mathbb{Z}_m$-symmetric state of affairs where time reversal changes this number and yet brings it back to its original value when operated twice.

This demonstrates that a chain of  $\mathbb{Z}_m$ parafermions with $m$ odd has only one symmetry protected phase. As a result, when the cyclic group of the parafermionic chain $ \mathbb{Z}_N$ has no $ \mathbb{Z}_2$ subgroup (i.e., for $ N$ odd), in the presence of modified time reversal symmetry  ${\cal T}^2=1 $ each of its gapped phases splits into two distinct phases.

For $\nu\ne \pm1/m$, characterizing the topological properties of the state requires more than the filling factor. Abelian states are described by the integer-valued $K$-matrix and charge vector $t$. In particular, the degeneracy of the ground state on a torus (which is also the number of topologically distinct types of quasi-particles) is $\det K$, the filling factor is $t^TK^{-1}t$, and $e^*$, the charge of the smallest charge quasi-particle, is the minimum over all integer vectors $l$ of $l^TK^{-1}t$ (the charge is given in units of the electron charge).  The electron charge is always an integer multiple of $e^*$'s.

We consider a fractional topological insulator in which electrons with spin $\uparrow$ form a state characterized by $K$, and electrons with spin $\downarrow$ are in a state characterized by $-K$. When  $\frac{1}{e^*}=\det K\equiv N$, there are  $N$ types of quasi-particles may be created by combining $j=1\dots N$ quasi-particles of charge $e^*$.
Consequently, each of the different quasi particles has a different charge.
Furthermore, in that case $N$ is odd\cite{Note2}. Thus, independent of the precise way in which the edge is gapped in a $B$ region (there generally is more than one possible way), the domain operators will add or remove $\pm je^*$ dipoles to the edge. Similarly, in the superconducting regions all domain operators will add charges of $je^*$ to each spin direction. As a consequence, the interface consists of a $\mathbb{Z}_2$ parafermion and a $\mathbb{Z}_N$ parafermion.

Once the composition of the $ \mathbb{Z}_N$ parafermions is determined, the discussion above shows that in the presence of time reversal symmetry, each of the phases of the $ \mathbb{Z}_2\times Z_N$  parafermionic chain splits into four subgroups when time reversal symmetry is present.

When  $\det{K} e^*\neq 1$  there are  distinct quasi-particles carrying the same charge. For example, when $\det K e^*=2$, there are two topologically distinct quasi-particles of any charge $je^*$. Thus, several sets of dipoles may be created, and there may be different charge-conserving gapping mechanisms that condense different sets of dipoles. At the interfaces between regions gapped by different charge conserving mechanisms, neutral parafermionic modes are localized. This route allows for the construction of parafermionic interface modes in systems where charge is conserved.

In Sec. \ref{331} we analyse a simple example for this state of affairs, that arises in the $(331)$ quantum Hall state \cite{Halperin1983}.
The topological classification of the chains follows the same principle that it follows in the cases described above, where each topological phase splits into four subgroups in the presence of modified time reversal symmetry.

\subsubsection{Bosonic systems}\label{subs32}

One can consider a case where the system is composed of \emph{bosons}, rather than fermions. Fractional topological insulating states can be constructed in analogy with the fermionic case, by stacking fractional quantum Hall states of bosons with an opposite chirality. There is an important difference in the case where the fractional topological insulator is made of bosons, rather than fermions; this difference is easiest to exemplify on the $\nu=1/m$ case, with $m$ being even. In that case the boson number in the $S$ region on the edge is conserved only modulo one, rather than two in the fermionic case. This difference stems from the non-conservation of the number of bosons when the edge is coupled to a superfluid. As a consequence, the dimension of the low energy subspace is reduced from $2m$ per $S$ region to $m$ per $S$ region. In fact, it is further reduced to $m/2$ per $S$ region since terms that create and annihilate single bosons at interfaces are also able to measure the parity of the $e^*$ charges on the segment(where, as before, $e^*$ is the charge of the elementary quasiparticle, in units of the charge of a boson.). The interface parafermions are then $\mathbb{Z}_{m/2}$ parafermions \cite{Maghrebi2015}. Once this difference is taken into account, however, the classification of bosonic phases follow the same route as that of the fermionic ones.

\section{Fractionalization of symmetries} \label{fractionalization of symm}

\subsection{Review: Majorana chain}

We briefly review the results for the case of fermions with a time reversal symmetry $\mathcal{T}^2=1$.
Gapped  phases of Fermions in one dimension are classified by a $\mathbb{Z}_8 $ index. This classification follows from the fractionalization of the two symmetry operators: time reversal $T $ and fermion parity $ P$  in the low energy sector~\cite{Fidkowski2010,Turner2011,Fidkowski2011a}.
When projected on the low energy subspace, any symmetry operation can be represented as a product of two local operators that act near the left and right ends.
This allows us to define local time reversal and local fermion parity operators: $ \mathcal{T} = {\cal O}_L{\cal O}_RK$ and $ P=P_LP_R$. {Here, $P_R$, ${\cal O}_R$ ($P_L$, ${\cal O}_L$) are unitary operators with support near the right (left) ends of the system, respectively, and $K$ is complex conjugation.} While total time reversal $\mathcal{T}^2=1 $ commutes with total parity, $ [\mathcal{T},P]=0$, and $P$ is a bosonic operator, their local constituents obey:
\begin{enumerate}
\item  The local parity operator may be bosonic, $[P_L,P_R]=0 $ or fermionic, $\{P_L,P_R\}=0 $.
\item The local time reversal operator, {$\mathcal{T}_L \equiv{\cal O}_L K$,} may square to $\mathcal{T}_L^2 = \pm1 $.
\item The local time reversal operator may commute, $[P_L,\mathcal{T}_L]=0 $, or anticommute,  $\{P_L, \mathcal{T}_L\}=0$, with the local fermion parity operator.
\end{enumerate}
We therefore identify   $2^3=8 $ distinct gapped phases which can be distinguished by the local symmetry operations in the low energy subspace. These eight phases obey a $\mathbb{Z}_8$ group structure~\cite{{Fidkowski2010,Turner2011,Fidkowski2011a}}.

 \subsection{Parafermion chain}

We now turn to the case of a chain of parafermions with an even number, $L$, of zero modes $\chi_j$, that satisfy $\chi_i \chi_j = \chi_j \chi_i e^{\frac{2i \pi}{N} \mathrm{sgn}(i - j)}$, with $N \in \mathbb{Z}$ (i.e., $\mathbb{Z}_N$ parafermions).
Note that, compared to Eq.~(\ref{zeromodes1overm}), here we suppress the spin ($s$) index - for our purpose, it suffices to treat only parafermion operators of one spin species (e.g., up spin).

Phases of a one dimensional array of $\mathbb{Z}_N$ parafermion zero modes \emph{without} time reversal symmetry,  were classified by Motruk et al.~[\onlinecite{Motruk2013}]. It was found that if the factorization of $N$ into primes is of the form $N = p_1^{r_1} \dots p_l^{r_l}$, then the overall number of distinct phases is $N_{\mathrm{phases}} =  (r_1+1) \dots (r_l+1)$, equal to the number of subgroups of $\mathbb{Z}_N$.

Here we consider a chain of parafermionic zero modes with time reversal that satisfies
$\mathcal{T}^{2}=1$, and a $\mathbb{Z}_{N}$ symmetry that corresponds either
to the total charge, $\mathcal{Q}=e^{i\pi\hat{Q}}$, or spin, $\mathcal{S}=e^{i\pi\hat{S}}$.
We denote the $\mathbb{Z}_{N}$ symmetry by $\mathcal{U}=\mathcal{S}$ or $\mathcal{Q}$;
in either case, it satisfies $\mathcal{U}^{N}=1$. Physically, these symmetries arise from the conservation of the fractional part of the spin or charge on the edge; e.g., on an FTI edge coupled to a superconductor, the total charge is conserved $\mathrm{mod}(2e)$, the charge of a Cooper pair. Whether the $\mathbb{Z}_N$ symmetry of the parafermion chain corresponds to $\mathcal{Q}$ or $\mathcal{S}$ depends on the physical realization. If the parafermion modes are formed at the ends of S regions surrounded by a large B region, as in Fig.~\ref{system}, the Hamiltonian commutes with $\mathcal{Q}$, the total fractional charge of the S regions. In a ``dual'' system where the B and S regions are interchanged, the symmetry is $\mathcal{S}$, the total fractional spin of the B regions.

The charge (spin)
operator is even (odd) under time reversal, respectively; hence, the
total charge and spin satisfy the following relations with time reversal:
\begin{eqnarray}
\mathcal{Q}\mathcal{T} & = & \mathcal{T\mathcal{Q^{\dagger}}},\label{eq:Q}\\
\mathcal{S}\mathcal{T} & = & \mathcal{T\mathcal{S}}.\label{eq:S}
\end{eqnarray}
In the ground state subspace of a gapped phase, the symmetry operator
can be written as a product of quasi-local operators that act near
the left and right ends of the system: $\mathcal{U}=\mathcal{U}_{L}\mathcal{U}_{R}$.
These operators can have exchange statistics corresponding to any
number of parafermion zero modes operators:
\begin{equation}
\mathcal{U}_{L}\mathcal{U}_{R}=e^{i\frac{2\pi n}{N}}\mathcal{U}_{R}\mathcal{U}_{L}.\label{eq:Uex}
\end{equation}
Similarly, the time reversal operation can be ``fractionalized'' into
a product of quasi-local unitaries that act near the left or right
ends, times complex conjugation: $\mathcal{T}=\mathcal{O}_{L}\mathcal{O}_{R}K$.
We can define local time reversal operations according to $\mathcal{T}_{R,L}=\mathcal{O}_{R,L}K$.

The different possible phases of the system are distinguished by the
relations between the local symmetry operators; each phase corresponds
to a unique projective representation of the symmetry at the ends
(or ``symmetry fractionalization''). Eq. (\ref{eq:Uex}) is an example
for such a relation. For $N$ prime, there are two possible distinct
phases: $n=0$ and $n=1$. (In this case, any value $1<n<N$ is equivalent
to $n=1$, since one can always redefine the symmetry operator as
$\mathcal{U}\rightarrow\mathcal{U}^{n}$.) In addition, in the presence
of time reversal with $\mathcal{T}^{2}=1$, there is the possibility
that $\mathcal{T}_{R,L}^{2}=\pm1$, corresponding to distinct phases.
The phase with $\mathcal{T}_{R,L}^{2}=-1$ is analogous to the Haldane
phase of integer spin chains~\cite{Haldane1983a}, where there is a Kramers pair
of zero modes at every end of the system, even though the ``microscopic''
time reversal operator (that acts on a single site) does not support
a Kramers' degeneracy~\cite{Pollmann2012}. In analogy with the Majorana chain, could there
be additional non-trivial phases, resulting from a non-trivial exchange
relation between time reversal and $\mathcal{U}_{R,L}$?

To address this question, let us analyze separately the cases $\mathcal{U=Q}$
and $\mathcal{U=S},$ as follows:

\subsubsection{$\mathcal{U=Q}$}

Let us write the charge operator as $\mathcal{Q}=\mathcal{Q}_{L}\mathcal{Q}_{R}$.
We assume that
\begin{equation}
\mathcal{Q}_{L}\mathcal{T}=e^{i \alpha}\mathcal{T}\mathcal{Q}_{L}^{\dagger},\,\,\mathcal{Q}_{R}\mathcal{T}=e^{i\frac{2\pi n}{N}-i\alpha}\mathcal{T}\mathcal{Q}_{R}^{\dagger}.\label{eq:QT}
\end{equation}
Here, $\alpha$ is a phase that we aim to find. The second equality follows from the requirement of Eq. (\ref{eq:Q}).
In order to fix the overall phase of $\mathcal{Q}_{R,L}$, we note
that
\begin{eqnarray}
\mathcal{Q}^{N} & = & \left(\mathcal{Q}_{L}\mathcal{Q}_{R}\right)^{N}\nonumber \\
 & = & e^{i\frac{2\pi n}{N}\frac{N(N-1)}{2}}\mathcal{Q}_{L}^{N}\mathcal{Q}_{R}^{N}=1.\label{eq:QN}
\end{eqnarray}
Hence, we can choose $\mathcal{Q}_{L}^{N}=1$, $\left(e^{i\frac{\pi n(N-1)}{N}}\mathcal{Q}_{R}\right)^{N}=1$.
By commuting $\mathcal{T}$ with $\mathcal{Q}_{L}^{N}=1$, we get
{[}using Eq. (\ref{eq:QT}){]}
\begin{equation}
\mathcal{Q}_{L}^{N}\mathcal{T}=e^{iN\alpha}\mathcal{T}\left(\mathcal{Q}_{L}^{\dagger}\right)^{N}.\label{eq:QN1}
\end{equation}
Hence, we conclude that $e^{iN\alpha}=1$. We write $\alpha=\frac{2\pi k}{N}$,
where $0\le k<N$.

Now, let us examine the consequences of Eq. (\ref{eq:QT}). In particular,
consider an eigenstate of $\mathcal{Q}_{L}$, $\vert q_{L}\rangle$,
such that $\mathcal{Q}_{L}\vert q_{L}\rangle=e^{i\frac{2\pi q_{L}}{N}}\vert q_{L}\rangle$
(where $0\le q_{L}<N$). Then,
\begin{eqnarray}
\mathcal{Q}_{L}\mathcal{T}\vert q_{L}\rangle & = & e^{i\alpha}\mathcal{T}\mathcal{Q}_{L}^{\dagger}\vert q_{L}\rangle=e^{i\alpha}\mathcal{T}e^{-i\frac{2\pi q_{L}}{N}}\vert q_{L}\rangle\nonumber \\
 & = & e^{i\frac{2\pi\left(k+q_{L}\right)}{N}}\mathcal{T}\vert q_{L}\rangle.
\end{eqnarray}
Therefore, $\mathcal{T}\vert q_{L}\rangle\propto\vert q_{L}+k\rangle$.
Iterating this relation twice, we get
\begin{equation}
\mathcal{T}^{2}\vert q_{L}\rangle=\vert q_{L}+2k\rangle\propto\vert q_{L}\rangle.
\end{equation}
Therefore, we conclude that $2k=0\mathrm{mod}(N)$. If $N$ is even,
this leaves us with the possibilities of $k=0,\,N/2$; for $N$ odd,
only $k=0$ is possible.

\subsubsection{$\mathcal{U=S}$}

In this case, we can retrace the steps in Eq. (\ref{eq:QT}-\ref{eq:QN1})
for $\mathcal{S}_{L,R}$, obtaining
\begin{equation}
\mathcal{S}_{L}\mathcal{T}=e^{i\alpha}\mathcal{T}\mathcal{S}_{L},\,\,\,\mathcal{S}_{R}\mathcal{T}=e^{-i\alpha}\mathcal{T}\mathcal{S}_{R},\label{eq:ST}
\end{equation}
where $\alpha=\frac{2\pi k}{N}$, and $\left(\mathcal{S}_{L}\right)^{N}=1$.
We are free to redefine the left and right spin operators as $\mathcal{S}_{L}=e^{i\frac{2\pi p}{N}}\tilde{\mathcal{S}}_{L}$,
$\mathcal{S}_{R}=e^{-i\frac{2\pi p}{N}}\tilde{\mathcal{S}}_{R}$.
Inserting $\tilde{\mathcal{S}}_{L}$ into Eq. (\ref{eq:ST}), we find
that it has a modified exchange relation with $\mathcal{T}$:
\begin{equation}
\tilde{\mathcal{S}}_{L}\mathcal{T}=e^{i\frac{2\pi\left(k-2p\right)}{N}}\mathcal{T}\tilde{\mathcal{S}}_{L}.
\end{equation}
Hence, we should identify any two values of $k$ that differ by an
even integer as two gauge-related descriptions of the same phase.
If $N$ is odd, we can always choose $p$ such that $k-2p=0\mathrm{mod}(N)$.
For $N$ even, on the other hand, there are two distinct values of
$k$: $k=0$ or $1$, corresponding to two distinct phases.

Summarizing this section, we find that in the presence of time-reversal symmetry that squares to $+1$, the classification of the gapped phases of $\mathbb{Z}_N$ parafermion chains becomes enriched. If $N$ is odd, each one of the distinct phases in the absence of time reversal symmetry now comes in two different flavors, depending on whether the local time time reversal operation squares to $+1$ or $-1$. If $N$ is even, there is an additional binary distinction, depending on the relation between the local charge or spin operations and time reversal; this distinction corresponds to the phase $\alpha$ in Eqs.~(\ref{eq:QT}) or (\ref{eq:ST}) (corresponding to cases where the total fractional charge or spin of the chain is conserved, respectively) taking the value $0$ or $\pi/N$. Thus, in the even $N$ case, each of the distinct phases in the absence of time reversal ``splits'' into four distinct phases when time reversal symmetry is imposed.

\section{Construction of gapped phases}\label{construction}

Below, we exemplify the symmetry fractionalization arguments described in Sec.~\ref{fractionalization of symm} by constructing the different possible phases in different physical examples. 

We first review the gapped phases of $\mathbb{Z}_m$ parafermions that are distinct even without any additional symmetry.
For $m$ prime, there are two distinct phases that correspond to the two ways in which the parafermionic modes may dimerize in pairs. Coupling a pair of zero modes annihilates  the domain trapped between them  and connects the two external domains, thus reducing the number of both superconducting and backscattering domains by one. This reduces the overall degeneracy from $m^N $ for $ N$ superconducting  segments  to $m^{(N-1)}$.

The ``trivial'' phase is constructed by coupling pairs of parafermions across the superconducting segments (namely, by making the superconducting segments shorter).  In the open chain geometry portrayed in Fig.~\ref{system} the Hamiltonian then reads:
 \begin{align}
H_{z=0} = -t \sum_{j=1}^L \eta_{2j-1}^\dag\eta_{2j},
\end{align}
where $t>0$ is a coupling constant. The resulting system has a single  backscattering segment on the outer edge and a non degenerate ground state. The non trivial phase is constructed by making all backscattering segments shorter:
\begin{align}
H_{z=1} = -t \sum_{j=1}^L \eta_{2j}^\dag\eta_{2j+1},
\end{align}
leaving the  outer edge of the FTI divided into one backscattering segment and one superconducting segment. The two interfaces between the different domains  host a pair parafermionic zero modes.

\subsection{Topological phases of $ \mathbb{Z}_m$ parafermionic modes with $m$ odd and prime and  $ {\cal T}^2=1$}

In the presence of a modified time reversal symmetry $ {\cal T}^2=1$, a richer phase structure emerges, which depends on {the parity of $m$.} For concreteness we consider the case of filling fraction $ m=3$, 
although a similar construction can be generalized to other values of $m$.
Our system is composed of alternating S and  B domains, surrounded by a large S domain. The low-energy states are labelled by the values of the fractional spin in each B domain, that take the values $s = -2/3,0,2/3 $. We define the following pseudo-spin 1 operators that act on the B domains:
\begin{align}
\Sigma_j^z &= P_{2/3}^j-P_{-2/3}^j,\\
\Sigma_j^+\Sigma_{j+1}^- &=\left[1-P_{-2/3}^j\right]\left[1-P_{2/3}^{j+1}\right]e^{i2\pi \hat{Q}_{j+1}},
\end{align}
where:
\begin{align}
P_s^j =\frac{1}{3}\left[ 1+ e^{i2\pi (\hat{S}_j-s)}+e^{-i2\pi (\hat{S}_j-s)}\right].
\end{align}
Here $ e^{i\pi \hat{Q}_j}$, and $ e^{i\pi \hat{S}_j}$ are the local charge and spin operators in the S and B $j$th domain, respectively. We then construct the following Hamiltonian:
\begin{align}
H = \sum_i J_\bot \Sigma_i^+\Sigma_{i+1}^- +J_z \Sigma_i^z \Sigma_{i+1}^z
\end{align}
In this basis time reversal acts as ${\cal T}=e^{i\pi \Sigma_y}K $.
Since this Hamiltonian only involves the operators  $e^{2i\pi \hat{Q}_i} $ and $ e^{2i\pi \hat{S}_i}$, it commutes with the total charge/spin $[H,e^{2i\pi \hat{Q}_{\rm tot}}]=0$, and thus does not lift the global $\mathbb{Z}_3$ symmetry.
For small enough $J_z$ this model is in the Haldane phase \cite{Haldane1983a,Haldane1983b} which supports a pseudospin-$1/2$ degree of freedom at its two ends. When the spin projection $\Sigma_z $ is a good quantum number, this local Kramers' doublet    corresponds to  eigenvectors of the local spin operator with  eigenvalues $\pm1/3 $:
\begin{align}
 {\cal S}_L |s_L = \pm1/3\rangle= e^{\pm i\frac{\pi}{3} }|s_L = \pm1/3\rangle.
\label{spinhalf}
\end{align}

Is the composite Haldane phase constructed above  topologically equivalent to the standard Haldane phase of a spin $1$ chain ($k=4$)? To address this question we first note that the standard Haldane phase is its own inverse. This implies  that a pair of  chains in  the Haldane phase are topologically equivalent to a trivial phase, and that their corresponding edge states can be coupled to a spin singlet state without breaking time reversal symmetry. However, the local antiferromagnetic term that couples the two spin $1/2$ end states  into a non degenerate spin singlet, includes  `flip flop' terms which  change the  edge spin on each chain. For the case of the composite Haldane phase constructed above, flipping the state of the pseudospin doublet in Eq.~(\ref{spinhalf}) requires a transfer of a fractional spin of $2/3 $ which cannot be achieved by coupling the system to a standard spin $ 1$ chain.
Hence the  only allowed coupling terms between this composite Haldane phase and a standard Haldane phase are of the form $ s_z \Sigma_z$, which leave the ground state of the two chain system four fold degenerate.

There is however a way to circumvent this restriction. We may couple the Kramers' doublet that appears on the end of the composite Haldane phase to a single B segment with a 3-fold ground states degeneracy, $|S = 0,2/3, -2/3\rangle $.
The three ground states of the additional B segment are eigenvectors of the local spin operator:
\begin{align}
{\cal S}_L |S\rangle = e^{\pm i\frac{2\pi S}{3} }|S\rangle.
\end{align}
We couple the additional B segment to the pseudospin $1/2$ degree of freedom at the end through the local Hamiltonian:
 $H = -J S s_L$ with $J>0$.
The coupled system has two degenerate ground states  corresponding to $|\pm\rangle = |S=\pm 2/3 \rangle \otimes | s_L = \pm1/3\rangle$, at the left end of the open chain. One can readily verify that the two states form a Kramer's pair, and that they both have the same spin eigenvalue:
\begin{align}
{\cal S}_L |\pm\rangle = e^{ i\frac{\pi}{3}(2S+s_L) }|\pm\rangle = -|\pm\rangle.
\end{align}
The resulting coupled system is therefore the analog of the $k=4  $ Fermionic phase with $[{\cal T}_L,{\cal S}_L]=0$. In particular,
the edge states of this system can be gapped by coupling it to an ordinary spin $1$ chain in the Haldane phase.

\subsection{Topological phases of $ \mathbb{Z}_{m}$ parafermionic modes with $m$  even}
\subsubsection{Emergent fermions in bosonic systems}

We next construct the topological phases  of parafermionic modes at the edge of a \emph{bosonic} fractional topological insulator. We will focus on the case of a  $\nu = 1/m$ Laughlin state for ``spin up'' bosons and a $\nu = -1/m$ for ``spin down," with $m$ even. The considerations of Sec.~\ref{fractionalization of symm} suggest that in the presence of time reversal symmetry, each of the distinct gapped phases on the edge should split into four classes. Here, we construct explicit Hamiltonians that place the system in each of these phases. The new phases that appear in the presence of time reversal symmetry can be traced back to the emergence of zero modes with \emph{fermionic} self-statistics at the edge. The new phases at the edge of a bosonic FTI are analogous to the phases of a Majorana chain with time reversal symmetry (class BDI).

\begin{figure}[h]
\includegraphics[width=0.45 \textwidth]{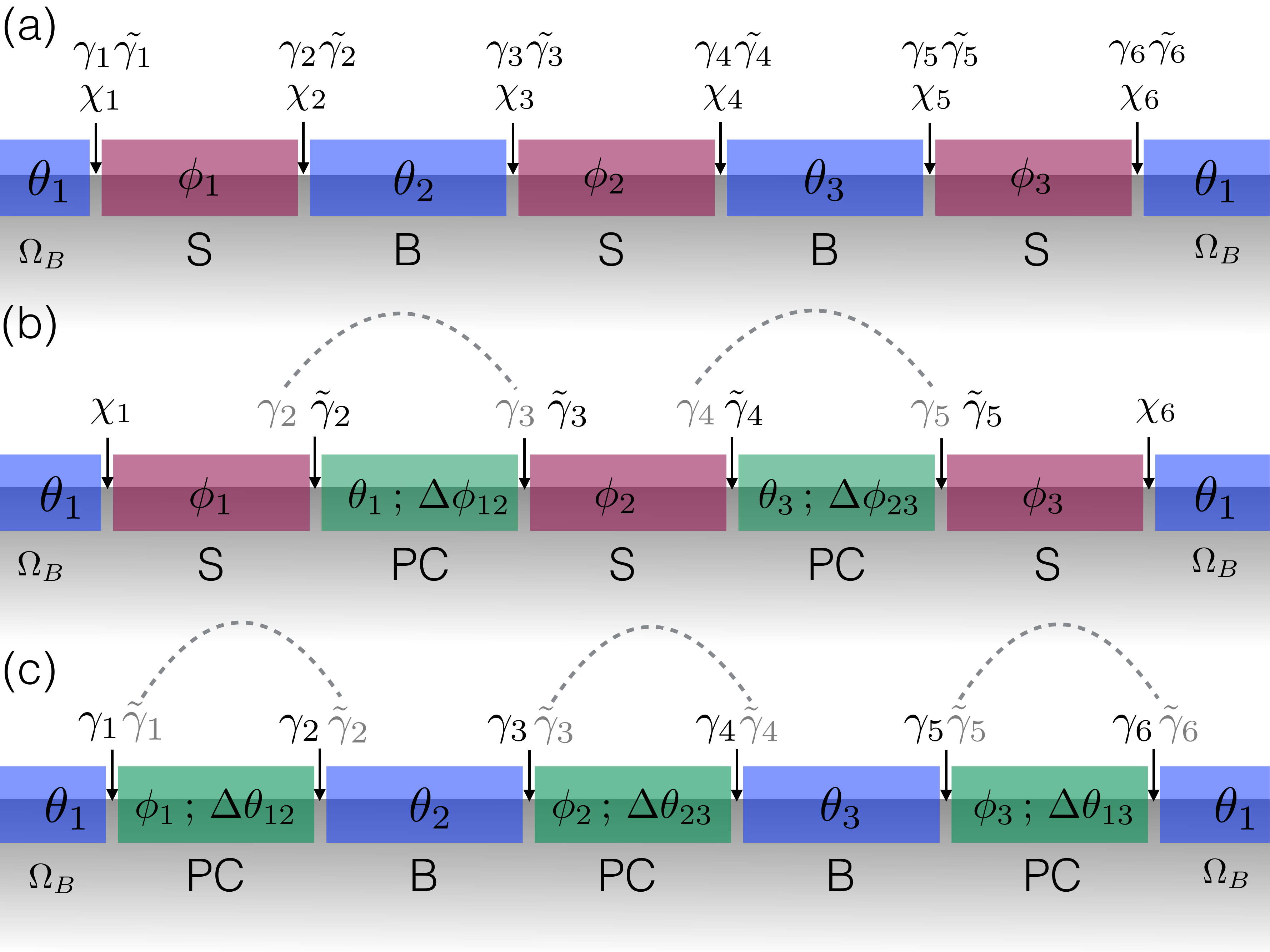}
\caption{(a) The edges of the $ \nu = 1/8$ fractional topological insulator can be gapped by alternating  segments of superfluid (S) and backscattering (B) regimes. The  interface between the two domain hosts a $\chi_j$ zero mode, which can be factorized into two non commuting  self fermionic operators, $\chi_j\sim \gamma_j\tilde{\gamma}_j$. (b) Coupling the $ \gamma_j$ modes across the B segments transforms the segment to a   parafermion condensate (PC).  The resulting system composed of alternating segments of S and PC realizes a Majorana chain in class BDI. (c) An alternative way to construct a Majorana chain can be achieved  by coupling   $ \tilde{\gamma}_n$ modes across the S segments. }
\label{chain}
\end{figure}

As in the fermionic $\nu = 1/m$ case, the edge states of a bosonic $\nu = 1/m$ FTI can be gapped either by backscattering between the edges, or by coupling the edges to a bosonic superfluid.
The pair tunneling and  backscattering potentials that gap the edge take the following respective forms:
\begin{align}
H_{S} &\propto  \psi_R^\dag \psi_L^\dag +{\rm h.c.} \sim \cos 2m \phi\\
H_{B} &\propto \psi_R^\dag \psi_L +{\rm h.c.} \sim \cos 2m \theta.
\end{align}
where  $ \psi_{R,L}^\dag \sim  e^{-im (\phi\pm \theta)} $ creates a right/left moving boson, and the bosonic fields satisfy the commutation relations
\begin{equation}
[\phi(x),\theta(x')]=\frac{i\pi}{m}\Theta(x'-x).
\end{equation}

In an edge with alternating backscattering and superfluid domains (see Fig.~\ref{chain}), either  $e^{i2\phi(x_j)} $ or $e^{i2\theta(x_j)}$ obtain a non-zero expectation value in the ground state manifold, where $x_j$ is the coordinate of a point inside the corresponding domain. The two  domain operators satisfy the exchange relations:
\begin{equation}
e^{2i\phi(x_{j})}e^{2i\theta(x_{j'})}=e^{4\frac{i\pi}{m}\Theta(x_{j}-x_{j'})}e^{2i\theta(x_{j'})}e^{2i\phi(x_{j})}.
\end{equation}

At the each interface between the two domains a zero mode appears. The zero modes satisfy a $\mathbb{Z}_{m/2} $ exchange rule:
\begin{eqnarray}
 \nonumber
 \chi_{2j-1} &=& e^{2i\phi(x_{j})+2i\theta(x_{j})}, \\
 \chi_{2j} &=& e^{2i\phi(x_{j})+2i\theta(x_{j+1})}.
 \end{eqnarray}
Hence, there must be $m/2$ degenerate ground states per superfluid (or backscattering) domain~\cite{Maghrebi2015}.

The simplest bosonic FTI state, $m=2$, has no ground state degeneracy at its edge. For $m=4$, there is a two-fold ground state degeneracy per superfluid domain, and the $\chi_j$ zero mode operators obey self-fermionic statistics: $\{ \chi_i, \chi_j \} = 0$. In this case, the classification of possible phases can be read off from the case of Majorana chains~\cite{Fidkowski2010, Turner2011, Fidkowski2011a}. There are two topologically distinct phases in the absence of time reversal symmetry, and an interface between them carries a single Majorana zero mode. In the presence of time reversal that satisfies ${\cal T}^2=1$, each of the two phases splits into four distinct phases~\cite{}, as expected from the considerations of Sec.~\ref{fractionalization of symm}.

We stress the difference between the phases at the edge of a $\nu=1/4$ bosonic FTI and the phases of an ordinary Majorana chain: although the mathematical description of these two systems is identical, their physical nature is different. For example, even though the $\chi_j$ zero mode operators at the edge of a bosonic FTI satisfy self-fermionic statistics, they are \emph{not} locally related to any microscopic degrees of freedom, which are bosonic in this problem. As such, they cannot be detected by tunneling particles from  outside the FTI system. Instead, the $\chi_j$ modes absorb or emit fractional quasi-particles of the bosonic FTI at zero energy. Their fermionic statistics is an emergent, rather than a microscopic, property.

We henceforth analyze the case $m=8$, which is a less trivial example that illustrates the general principle.
In the absence of any additional symmetries except the inherent $\mathbb{Z}_4$ symmetry, we can construct two topologically distinct phases corresponding to the two ways in which zero modes can be dimerized. For convenience, we define the ``vacuum'' or trivial phase as the phase in which the $\chi_i$ zero modes are coupled across the superconducting regions (see Fig.~\ref{chain}); we will also refer to this phase as the $B$ phase. The ``non trivial'' phase $S$, in which each pair of $\chi_i$ modes across a backscattering region are coupled, hosts a $\mathbb{Z}_4$ parafermionic zero mode at each interface with the trivial phase.

For the case of $m=8$ there is a third possible phase. To construct this phase, we note that we can write the $\mathbb{Z}_4$ zero mode operators in the following way:
\begin{eqnarray}
\nonumber
\chi_j &\sim & \tilde{\gamma}_j^\dag \gamma_j,
\end{eqnarray}
where
\begin{eqnarray}
\nonumber
 \gamma_{2j-1} =i e^{4i\phi(x_{j})+2i\theta(x_{j})},\ \   & \tilde{\gamma}_{2j-1}=i e^{2i\phi(x_{j})+4i\theta(x_{j})},  \\
\nonumber
 \gamma_{2j} =e^{4i\phi(x_{j})+2i\theta(x_{j+1})},\ \   & \tilde{\gamma}_{2j}=e^{2i\phi(x_{j})+4i\theta(x_{j+1})}.
\end{eqnarray}

While the two sets of operators $\gamma_j $ and $\tilde{\gamma}_j $ do not commute, each of them obeys fermionic self-statistics: $\{ \gamma_j, \gamma_{j'}\} =0$ and $\{\tilde{\gamma}_j, \tilde{\gamma}_{j'}\} = 0$.
Coupling  $ \gamma$ operators across a backscattering domain by turning on the following Hamiltonian:
\begin{eqnarray}
H_{2j} = i t \gamma_{2j}\gamma_{2j+1},
\end{eqnarray}
where $t$ is the coupling strength, commutes with the  $\tilde{\gamma}_{2j} $, $\tilde{\gamma}_{2j+1}$ zero modes at each interface and thus
leaves these two operators as zero modes (see Fig~\ref{chain}b). Alternatively, a perturbation that couples $ \tilde{\gamma}$  across a superconducting domain
\begin{eqnarray}
H_{2j-1} = it \tilde{\gamma}_{2j-1}\tilde{\gamma}_{2j}
\end{eqnarray}
commutes with the remaining  $ \gamma$ zero modes, see Fig \ref{chain}c.
In either cases,  coupling the self-fermionic modes across the respective domain results in a gapped segment in which  the commuting operators $e^{4i \phi} $ and $e^{4i \theta} $ acquire a non-zero expectation value in the ground state manifold (i.e., these operators ``condense''). This phase has been dubbed a parafermion condensate phase (PC)~\cite{Motruk2013}, as it is characterized by the parafermion operator $\chi^2 \sim e^{4i\theta + 4i\phi}$ having a non-zero expectation value everywhere in the bulk. (Note that $\chi^2$ is a self-boson, and hence it can condense).

Consequently, in the absence of any additional symmetries we find three  distinct gapped phases of a $ \mathbb{Z}_4$ parafermionic chain, and correspondingly  three boundary  zero modes at interfaces between them:
\begin{eqnarray}
\nonumber
\chi_{\rm B-S}&=&\gamma \tilde{\gamma}, \\
\nonumber
\chi_{\rm B-PC} &=&\gamma, \\
\chi_{\rm S-PC} &=&\tilde{\gamma}.
\end{eqnarray}

We next turn to the classification of these systems with time reversal symmetry.
Based on the general  arguments presented above we expect each phase to split into 4 distinct phases. Below we will construct these phases explicitly. We start from  an open chain of zero modes formed  by an array of  alternating  backscattering and parafermion condensate  domains (Fig. \ref{chain}), with one long backscattering domain acting as our vacuum. Each domain wall between B and PC has a zero mode $\chi_{\rm B-PC}=\gamma$.
We define a parity operator
\begin{eqnarray}
P = -i\gamma_{2j-1}^\dag\gamma_{2j}= e^{2i(\theta(x_{j+1})-\theta(x_{j}))}.
\end{eqnarray}
Since $ [{\cal T},P]=0$, the two zero modes transform differently under time reversal: ${\cal T}^{-1} \gamma_{2j-1}{\cal T}=\gamma_{2j-1} $ and ${\cal T}^{-1} \gamma_{2j}{\cal T}=-\gamma_{2j} $ (where we have fixed an overall phase of the operator to $ \gamma_{2j-1}^2 =\gamma_{2j}^2 =1$).
It follows that chain of $\gamma$ zero modes constructed by alternating segments of B and PC realizes a  Majorana chain in class BDI.

 By coupling the $\gamma_n$  Majorana-like modes to each other, we can realize 8 distinct phases, with $N_\gamma = 0...7$ protected zero modes at end of the open chain. We denote the phases with an even number of $\gamma $ zero modes at each end, $B_0,B_2,B_4,B_6 $ and the phases with odd number of zero modes $C_1,C_3,C_5,C_7 $,  where the subscript counts the number of Majorana-like zero modes at each end of the open chain.
Classifying the phases as $ B_i$ or $ C_i$ follows from the effect of time reversal symmetry breaking terms,  where the boundary between phases with an even number of Majorana modes and the vacuum (the large B region) can be gapped by breaking time reversal resulting in a trivial phase, while phases with an odd number of Majorana modes at each end cannot be fully gapped, and would correspond to a PC  wire  with a single Majorana-like mode at each end.

An alternative way to realize a  Majorana chain is by constructing alternating segments of S and PC (Fig. \ref{chain}) with one long backscattering domain acting as our vacuum.
For concreteness we consider the outer segments to be S domains. Each interface between  S and PC segments  hosts a $\tilde{\gamma} $ zero mode
obeying fermionic self statistics, while the boundary domain walls between the outer S domains and the long B segment host a $ \chi_{\rm B-S}$
 mode.  The construction of a parity operator
 $ P = -i\tilde{\gamma}_{2j}^\dag\tilde{\gamma}_{2j+1}$
 readily follows the case of the $\gamma $ chain. By coupling the $\tilde{\gamma}_n$  Majorana-like modes to each other, we can realize 8 distinct phases, with a single $\chi $ zero mode and $N_{\tilde{\gamma}} = 0...7$ protected  Majorana-like modes at end of the open chain. We denote the phases with an even number of $\tilde{\gamma}$ zero modes at each end, $S_0,S_2,S_4,S_6 $ and the phases with odd number of zero modes $\tilde{C}_1,\tilde{C}_3,\tilde{C}_5,\tilde{C}_7 $.
 An even number of $ \tilde{\gamma}$ can be gapped by a time reversal symmetry breaking perturbation, resulting in a S chain with a single $\chi $ zero mode at each end.

\begin{figure}[h]
\includegraphics[width=0.47 \textwidth]{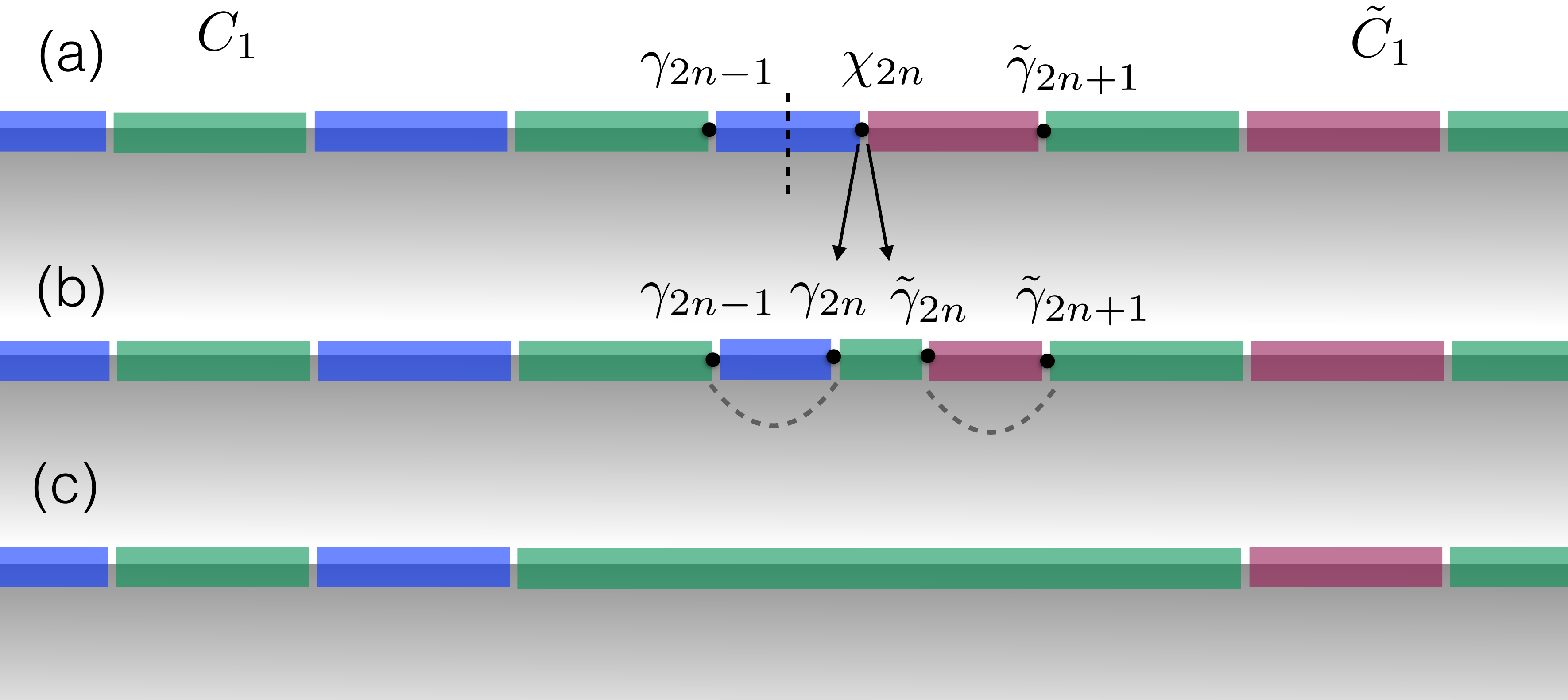}
\caption{ Local perturbation can gap up the zero modes that appear at the interface between $ C_1$ and $ \tilde{C}_1$ segments. (a) The dashed line marks the interface between the $ C_1$ phase with a single $ \gamma_{2n-1}$ at its end, and  the $ \tilde{C}_1$ phase with one $\chi_{2n} $ and one $ \tilde{\gamma}_{2n+1}$ at its end. (b) The $\chi_{2n} $ mode can be split into two Majorana like modes: $\gamma_{2n} $ and $\tilde{\gamma}_{2n} $ by nucleating a narrow PC region in between. (c) Coupling each pair of $ \gamma$ ($\tilde{\gamma} $) modes across the B (S) domain gap up the interface zero modes without violating time reversal symmetry.}
\label{C1}
\end{figure}

We will next show that the  $\tilde{C}_i $ phases are topologically equivalent to the $C_i $ phases. In particular, this requires that the zero modes that appear at the interface between the two phases can be gapped  by a local perturbation without breaking  time reversal symmetry. We will demonstrate this explicitly for the case of  an interface between $ C_1$ and $\tilde{C}_1$, see Fig. \ref{C1}. A similar construction can be made for the other three $ C_i$ classes.
The $\chi_{2n} $ mode at the interface can be split into two Majorana-like modes: $\gamma_{2n} $ and $\tilde{\gamma}_{2n} $, by nucleating a narrow PC region at the interface between the B and S segments. The resulting boundary system had $N_{\gamma}=2 $, $N_{\tilde{\gamma}}=2 $ and $ N_{\chi}=0$.  We then note that the pair of $\gamma$ modes located at the two ends of the B segment transform differently under time reversal operation. Hence by coupling the $ \gamma$ modes across the B segment:
\begin{eqnarray}
\delta H_\gamma = i t \gamma_{2n}\gamma_{2n-1}
\end{eqnarray}
we can lift the degeneracy associated with them, without violating time reversal symmetry. Similarly, the following time reversal symmetric term:
\begin{eqnarray}
\delta H_{\tilde{\gamma}} = i\tilde{\gamma}_{2n}\tilde{\gamma}_{2n+1},
\end{eqnarray}
 lifts the degeneracy associated with the S segment. These local perturbations gap up all interface zero modes while leaving the time reversal symmetry intact. This indicates that the  two phases are topologically equivalent.

We therefore conclude in the absence of any additional symmetries, a chain of  $ \mathbb{Z}_4$ parafermions realises 3 topological phases: B, S, PC. When the system is time reversal symmetric each phase splits into four, resulting in a total of 12 distinct gapped phases: $B_0,B_2,B_4,B_6 $, $S_0,S_2,S_4,S_6 $ and $C_1,C_3,C_5,C_7 $.  According to the arguments of Sec.~\ref{fractionalization of symm}, we expect each of the three phases that exist in the absence of time reversal to split into four distinct phases when time reversal is imposed. Thus, the 12 phases exhaust the list of possible phases with time reversal symmetry.

\subsubsection{Parafermions with no  superconductors}\label{331}
In this subsection we shows that ${\cal T}^2=1$ time-reversal-symmetric parafermion chains may be constructed even in a setting in which charge is conserved, i.,e., without resorting to coupling the edges to superconductors. This  may occur in fractional systems where the smallest fractional charge $ e^*$  and the degeneracy  (given by the determinant of the $K$ matrix) satisfy $e^*\det K \neq 1 $. In this case, there are several different quasi-particles that carry the same charge. It is then possible to have different charge conserving  mechanisms that condense topologically distinct sets of dipoles. The interfaces between domains with  topologically distinct gaps host neutral parafermionic modes.  We focus on a particular  example where the neutral zero modes are self fermions that  form a $\mathbb{Z}_2$ group.

We consider the  $(331)$ quantum Hall state, where the edge theory is governed by the Lagrangian~\cite{Wen1990,Lee1991,Levin2012}:
\begin{align}
\nonumber
{\cal L} &= \frac{1}{4\pi}\left(\partial_t \Phi^TK \partial_x\Phi-\partial_x \Phi^TV\partial_x\Phi\right)\\
&+\frac{1}{2\pi}\epsilon^{\mu\nu}t^T\partial_\mu\Phi A_\nu.
\end{align}
Here $ V$ is a velocity matrix, $ t=(1,1,1,1)^T$ is the charge vector, and the $K$-matrix  is given by:
\begin{align}
K= \left(
\begin{array}{ccc}
  \kappa & 0    \\
  0&-\kappa
\end{array}
\right)
\end{align}
with
\begin{align}
\kappa =\left(
\begin{array}{ccc}
3  & 1    \\
  1& 3
\end{array}
\right).
\end{align}
The bosonic fields satisfy the commutation relation:
\begin{align}
[\partial_x\Phi_i(x),\Phi_j(x')]=2\pi i \delta(x-x')K^{-1}_{ij}
\end{align}
Local operators composed of products of electron creation and annihilation operators take the form: $e^{i\Lambda^TK\Phi} $, where $\Lambda $ is a vector of integers (this follows from the requirement that their mutual statistics with any quasi-particle must be trivial). Charge conserving local operators satisfy $\Lambda^T t =0$. In accordance with Ref.~\onlinecite{Levin2013}, The edge states can be gapped by two topologically distinct charge conserving perturbations corresponding to:
\begin{align}
H_{1} &= \lambda_{1a}\cos\left(\Lambda_{1a}^TK\Phi\right)+ \lambda_{1b}\cos\left(\Lambda_{1b}^TK\Phi\right)\\
H_{2} &= \lambda_{2a}\cos\left(\Lambda_{2a}^TK\Phi\right)+ \lambda_{2b}\cos\left(\Lambda_{2b}^TK\Phi\right)
\end{align}
where $D=1,2 $ are the two domains, and $\Lambda_{1a}^T = (1,0,1,0)$,  $\Lambda_{1b}^T = (0,1,0,1)$, $\Lambda_{2a}^T = (0,1,1,0)$ $\Lambda_{2a}^T = (1,0,0,1)$.

Each domain has two  domain operators $ e^{i\Lambda_{a,b}^T\Phi} $, which  commute with the respective Hamiltonian.
Using products of domain operators in the two neighboring domains, we can construct  the following domain wall operator:
\begin{eqnarray}
\chi_n&=&e^{i(1,0,1,0)\Phi_n-i(0,1,1,0)\Phi_{n+1}}
\end{eqnarray}
(Different  choices  of pairs of domain operators  result in operators which are  related by a neutral boson).
These charge neutral operators satisfy fermionic exchange statistics with respect to each other:
\begin{equation}
\{\chi_{n},\chi_{n'}\}=2\delta_{n,n'},
\end{equation}
and thus behave as Majorana fermions.
The construction of gapped phases follows the usual one dimensional BDI fermionic chain giving rise to $8 $ classes in the presence of  an effective time reversal symmetry.

Note that, as in the bosonic case described above, although the $\chi_n$ operators are formally similar to ordinary Majorana zero modes, they have different physical properties. E.g., a physical electron \emph{cannot} be absorbed at zero energy at an interface between the two types of domains. Nevertheless, the structure of the possible phases at the edge in the presence of time reversal is the same as that of an ordinary Majorana chain.

\section{Conclusion}\label{conclusion}
We studied gapped phases of $\mathbb{Z}_N$  parafermionic chains in the presence of a modified time reversal symmetry ${\cal T}^2 =1 $.
Without time reversal, there are $N_{\rm phases} = (r_1+1)...(r_k+1)$ distinct phases, where $N=p_1^{r_1}\dots p_k^{r_k}$ is the decomposition of $N$ to prime factors. When time reversal symmetry is imposed, the resulting phase structure depends on the parity of $N$. 
If $N $ is odd, each of the  phases  splits into two  subclasses. This is since the class with no parafermionic end states  splits into a trivial phase and a symmetry protected Haldane phase. Conversely, when $N $ is even, each phase splits into four subclasses. The origin of this split is in the emergent Majorana fermions associated with even values of $N$. Akin to the classification of Majorana chains,  the trivial class then splits into four symmetry protected subclasses: the  trivial subclass, two anomalous subclasses with two Majorana modes at each end, and the Haldane phase which corresponds to  four Majorana modes at each end of the chain. We demonstrate the appearance of such emergent Majorana zero modes in a system where the constituents particles are  bosons as well as in the absence of any form of pairing.

D. M. acknowledges  support from the  Israel Science Foundation (Grant No. $ 737/14$) and
from the European Union's Seventh Framework Programme (FP7/2007-2013) under Grant No. 631064. EB acknowledges support from a Marie Curie CIG grant and from the European Research Council (ERC) under the European Union Horizon 2020 research and innovation programme (No. 639172). A.S. acknowledges support from the European
Research Council under the European Unions
Seventh Framework Program (FP7/2007-2013) / ERC
Project MUNATOP, the DFG (CRC/Transregio 183, EI
519/7-1), Minerva foundation, the U.S.-Israel BSF and Microsoft's Station Q.

\end{document}